\documentclass[preprint,aps,prd,showpacs,nofootinbib,tightenlines]{revtex4}

\usepackage{amsfonts}
\usepackage{multirow}
\usepackage{mathrsfs}
\usepackage{graphicx}
\usepackage{amsmath}
\usepackage{amssymb}
\usepackage{bm}
\usepackage{bbm}
\usepackage{slashbox}
\usepackage{color}

\def\bfnabla{\mbox{\boldmath $\nabla$}}

\def\bfsigma{\mbox{\boldmath $\sigma$}}

\def\em{{\rm em}}

\def\als{\alpha_{\rm s}}

\def\siml{{\ \lower-1.2pt\vbox{\hbox{\rlap{$<$}\lower6pt\vbox{\hbox{$\sim$}}}}\ }}
\def\simg{{\ \lower-1.2pt\vbox{\hbox{\rlap{$>$}\lower6pt\vbox{\hbox{$\sim$}}}}\ }}

\def\vbfD{{\ \lower-8pt\vbox{\hbox{\rlap{$\!\leftrightarrow$}\lower8pt\vbox{\hbox{$\!\bf D$}}}}\ }}
\def\dsl{\,\raise.15ex\hbox{/}\mkern-13.5mu D}

\def\OMIT#1{}

\newcommand{\nn}{\nonumber}

\newcommand{\beq}{\begin{equation}}
\newcommand{\eeq}{\end{equation}}
\newcommand{\bqa}{\begin{eqnarray}}
\newcommand{\eqa}{\end{eqnarray}}

\begin{document}
\title{\mbox{}\\[10pt]
Is the $J^P=2^-$ assignment for the $X(3872)$ compatible with the
radiative transition data?}

\author{Yu Jia}
\affiliation{Institute of High Energy Physics, Chinese Academy of
Sciences, Beijing 100049, China\vspace{0.2cm}}
\affiliation{Theoretical Physics Center for Science Facilities,
Chinese Academy of Sciences, Beijing 100049, China\vspace{0.2cm}}

\author{Wen-Long Sang}
\affiliation{Institute of High Energy Physics, Chinese Academy of
Sciences, Beijing 100049, China\vspace{0.2cm}}
\affiliation{Theoretical Physics Center for Science Facilities,
Chinese Academy of Sciences, Beijing 100049, China\vspace{0.2cm}}

\author{Jia Xu}
\affiliation{Institute of High Energy Physics, Chinese Academy of
Sciences, Beijing 100049, China\vspace{0.2cm}}
\affiliation{Theoretical Physics Center for Science Facilities,
Chinese Academy of Sciences, Beijing 100049, China\vspace{0.2cm}}

\date{\today}
\begin{abstract}
The very recent analysis by \textsc{BaBar} Collaboration indicates
that the $X(3872)$ may favor the quantum number $J^{PC}=2^{-+}$
rather than the previously assumed $1^{++}$. By pretending the
$\eta_{c2}(1D)$ charmonium to be the $X(3872)$, we study the
parity-even radiative transition processes $\eta_{c2}(1D)\to
J/\psi(\psi')+\gamma$ within several phenomenological potential
models. We take the ${}^3D_1$ admixture in $\psi'$ into account, and
consider the contributions from the magnetic dipole ($M1$), electric
quadrupole ($E2$), and magnetic octupole ($M3$) amplitudes. It is
found that the ratio of the branching fractions of these two
channels, as well as the absolute branching fraction of
$\eta_{c2}\to \psi'\gamma$, are in stark contradiction with the
existing \textsc{BaBar} measurements. This may indicate that the
$2^{-+}$ assignment for the $X(3872)$ is highly problematic.
\end{abstract}

\maketitle

The rising of a dozen of new charmonium resonances in recent years,
most of which are above the $D\overline{D}$ mass threshold, has
greatly reinvigorated the field of hadron spectroscopy. Among them,
the $X(3872)$ particle, perhaps being the most intensively studied
one from both theory and experiment, has occupied the central
stage~\cite{Brambilla:2004wf}. The $X(3872)$ was first discovered in
2003 by \textsc{Belle} Collaboration in the decay $B^+\to
J/\psi\pi^+\pi^-K^+$~\cite{Choi:2003ue}, subsequently confirmed by
\textsc{Babar}~\cite{Aubert:2004ns}, as well as
CDFII~\cite{Acosta:2003zx}, D0~\cite{Abazov:2004kp} in inclusive
$p\bar{p}$ collision experiments. Several unusual properties of this
particle, {\it i.e.} its mass in extreme proximity to the
$D^0\overline{D}^{*0}$ threshold, the rather narrow width, and the
large isospin violation seen in its decay pattern, have persuaded
many authors to believe that, rather than being a conventional
charmonium~\cite{Barnes:2003vb}, the $X(3872)$ may be of exotic
nature, {\it e.g.}, a loosely-bound $D^0$-$\overline{D}^{*0}$
molecule~\cite{Tornqvist:2003na}, or a diquark-antidiquark
cluster~\cite{Maiani:2004vq}, or a hybrid~\cite{Li:2004sta}.

Aside from its mass, the most important property about the $X(3872)$
is its  $J^{PC}$ quantum number. The even $C$-parity of the
$X(3872)$ has been firmly established since the recent observations
of its decay to $J/\psi\rho^0$~\cite{Abulencia:2005zc} and to $J/\psi\gamma$~\cite{Aubert:2006aj}.
By analyzing the decay angular
distribution in the process $X(3872)\to J/\psi\pi^+\pi^-$, CDF
Collaboration narrowed the possibilities of the $J^P$ down to $1^+$
or $2^-$~\cite{Abulencia:2006ma}. The former assignment seems more
appealing from the theoretical perspective, which is particularly
congenial to a $S$-wave $D^0 \overline{D}^{*0}$ molecule
interpretation. Although the smoking gun from the experimental side
is still not available yet, the $1^{++}$ assignment for the
$X(3872)$ has already been tacitly accepted in most of the recent
phenomenological works.

However, very recently there comes a quite unexpected, and, perhaps
disquieting, news from \textsc{BaBar} Collaboration. The latest
analysis of the decay $B\to J/\psi \omega K$ by \textsc{BaBar}
indicates that the $P$-wave orbital angular momentum for the $J/\psi
\omega$ system is more favored than the $S$-wave, which implies that
the $X(3872)$ may favor $J^{PC}=2^{-+}$ instead of the
universally-believed $1^{++}$~\cite{delAmoSanchez:2010jr}.
Therefore, we are compelled to re-scrutinize the properties of the
$X(3872)$. If future experiments will confirm the result of
\cite{delAmoSanchez:2010jr}, our perception on the nature of this
particle would have to be profoundly changed.

In light of the latest \textsc{BaBar}
analysis~\cite{delAmoSanchez:2010jr}, the most natural candidate for
$X(3872)$ would be the $\eta_{c2}(1D)$ meson. This $D$-wave
spin-singlet (denoted by the spectroscopic symbol ${}^1D_2$)
charmonium has been extensively studied in quark potential models
for decades. Its predicted mass is scattered in the range 3760-3840
MeV~\cite{Barnes:2003vb}. The width of $\eta_{c2}$ is believed to be narrow,  since the
decay into $D\overline{D}$ is forbidden by parity, and the energy
conservation does not allow it to disintegrate into
$D\overline{D}^*$. It can only decay through the strong and
electromagnetic transitions, exemplified by $\eta_{c2}\to \eta_c\pi\pi$ and
$\eta_{c2}\to h_c\gamma$, as well as through the OZI-forbidden
annihilation $\eta_{c2}\to gg$.
Each of these processes is expected to have a partial width of a few hundred keV.

One of the strongest objections to identifying the $X(3872)$ with
the $\eta_{c2}$ charmonium probably comes from the electromagnetic
transitions $\eta_{c2}\to
J/\psi(\psi')+\gamma$~\cite{Godfrey:2008nc}. Such
parity-conserving transitions flip the quark spin and change the
orbital angular momentum by two units, so there must be strong
multipole suppression, and one would expect a rather small branching
fraction for such decay processes. Thus, one is puzzled by the fact
why the \textsc{BaBar} Collaborations were able to observe these
radiative decay channels several years ago~\cite{Aubert:2006aj},
with only a limited statistics of the $X(3872)$ samples?

The motif of this paper is to quantify this objection, by presenting
a detailed study for the radiative transition processes
$\eta_{c2}\to J/\psi(\psi')+\gamma$. We will employ potential nonrelativistic QCD
(pNRQCD) as a convenient calculational device, and work with several
phenomenological potential models, as well as take the ${}^3S_1\!-\!{}^3D_1$
admixture effect for $\psi'$ into account. We will also identify the
contributions from several different multipoles, {\it e.g.}, the
magnetic dipole ($M1$), electric quadrupole ($E2$), and magnetic
octupole ($M3$) amplitudes.

Our key finding is that, no matter the mixing effect for $\psi'$ is
taken into account or not, the predicted branching ratio for
$\eta_{c2}\to \psi'+ \gamma$ is orders of magnitude smaller than the
lower bound that can be inferred from the \textsc{BaBar}
measurement~\cite{Aubert:2008rn}. Moreover, the ratio of the
branching fraction of $\eta_{c2}\to \psi'+ \gamma$ to that of
$\eta_{c2}\to J/\psi+ \gamma$ is orders of magnitude smaller than
the corresponding \textsc{BaBar} measurement for the $X$
particle~\cite{Aubert:2008rn}. These qualitative conclusions do not
vary with specific potential model. We thus tend to conclude that,
if the \textsc{BaBar} experiment is correct, the $\eta_{c2}$
assignment for the $X(3872)$ particle would become highly unlikely.
We hope that future higher-statistics experiments will help to
clarify the situation.

Before launching into the calculation, we first recall some related experimental
facts. \textsc{BaBar} has recently measured the following products of
two branching fractions~\cite{Aubert:2008rn}:
\begin{subequations}
\bqa
&& {\cal B}[B^\pm \to X(3872)K^\pm]\,{\cal B} [X(3872)\to J/\psi+
\gamma] = (2.8\pm 0.8\pm 0.1)\times 10^{-6},
\label{Babar:measure:jpsi:gamma}
\\
&& {\cal B}[B^\pm \to X(3872)K^\pm]\, {\cal B} [X(3872)\to \psi'+
\gamma] = (9.5\pm 2.7\pm 0.6)\times 10^{-6},
\label{Babar:measure:psiprime:gamma}
\eqa
\label{Babar:meas:jpsi:psi2s:gamma}
\end{subequations}
with $3.6\sigma$ and $3.5\sigma$ significance, respectively.
Therefore one can deduce the ratio of the branching fraction of
$X\to \psi'\gamma$ to that of $X \to J/\psi\gamma$:
\bqa
{{\cal B}[X(3872)\to \psi'+ \gamma] \over {\cal B}[X(3872)\to
J/\psi+\gamma]}={3.4\pm 1.4}.
\label{ratio:two:branch:ratio:Babar}
\eqa
This measurement is in serious conflict with the predictions made
from some specific $D^0\overline{D}^{*0}$-molecule
models~\cite{Tornqvist:2003na}, but the calculational framework
underlying those models may be questionable. Interestingly,
this ratio seems roughly compatible with the canonical $\chi_{c1}(2P)$
interpretation of the $X(3872)$, which decays to $J/\psi(\psi')+\gamma$
through the dominant electric dipole ($E1$)
transition~\cite{Barnes:2003vb,DeFazio:2008xq,Li:2009zu,Wang:2010ej}~\footnote{It seems that
the ${}^3S_1\!-\!{}^3D_1$ mixing effect has not been incorporated 
in most phenomenological analysis of $\chi_{c1}(2P)\to \psi^\prime\gamma$.}.

Very recently \textsc{Belle} Collaborations have also analyzed these
two radiative transition channels~\cite{Bhardwaj:2010}.
Their preliminary results are
\begin{subequations}
\bqa
&& {\cal B}[B^\pm \to X(3872)K^\pm]\,{\cal B} [X(3872)\to J/\psi+
\gamma] = (1.78^{+0.48}_{-0.44} \pm 0.12)\times 10^{-6},
\\
&& {\cal B}[B^\pm \to X(3872)K^\pm]\, {\cal B} [X(3872)\to \psi'+
\gamma] < 3.4\times 10^{-6}.
\eqa
\label{Belle:meas:jpsi:psi2s:gamma}
\end{subequations}
Their first measurement is consistent with \textsc{BaBar}'s result,
(\ref{Babar:measure:jpsi:gamma}). But \textsc{Belle} has not
observed any  $B^\pm \to \psi'\gamma K^\pm$ signals. Consequently,
\textsc{Belle} is only able to place an upper bound on the ratio
of these two branching fractions:
\bqa
{{\cal B}[X(3872)\to \psi'+ \gamma] \over {\cal B}[X(3872)\to
J/\psi+\gamma]} < 2.1,
\label{ratio:two:branch:ratio:Belle}
\eqa
at 90\% confidence level.

Thus far, our knowledge is only limited to the product of two
branching fractions, and it will be certainly useful to know the
absolute branching fraction of $X\to J/\psi(\psi')\gamma$. Using a
missing mass technique, \textsc{BaBar} was able to set an upper
bound for the absolute branching fraction of $B^\pm \to
X(3872)K^\pm$ some time ago~\cite{Aubert:2005vi}:
\bqa
&& {\cal B}[B^\pm \to X(3872)K^\pm]< 3.2 \times 10^{-4},
\label{absolute:br:B:X:K}
\eqa
at 90\% confidence level.

Combining the results of (\ref{Babar:meas:jpsi:psi2s:gamma}),
(\ref{Belle:meas:jpsi:psi2s:gamma}) together with
(\ref{absolute:br:B:X:K}), we are able to place some lower bounds on
the absolute branching fractions of the $X(3872)$ decays to
$J/\psi(\psi')+\gamma$:
\begin{subequations}
\bqa
&& {\cal B} [X(3872)\to J/\psi+ \gamma] > 5.9\times
10^{-3},\quad{\cal B}[X(3872)\to \psi'+ \gamma] > 1.9\times
10^{-2},\quad\textsc{Babar}
\nn\\
\\
&& {\cal B} [X(3872)\to J/\psi+
\gamma] > 3.8\times 10^{-3}.\quad \quad\quad\textsc{Belle}
\eqa
\label{inequality:X:jpsi:psi2S:gamma:exp}
\end{subequations}
We are unable to establish a meaningful inequality for $X\to \psi'+
\gamma$ from the latest \textsc{Belle} data.

Our central task then is to explicitly examine whether the decays
$\eta_{c2}\to J/\psi(\psi')+\gamma$ are consistent with the
experimental constraints listed in
(\ref{ratio:two:branch:ratio:Babar}),
(\ref{ratio:two:branch:ratio:Belle}), and
(\ref{inequality:X:jpsi:psi2S:gamma:exp}).

In passing, we note that there have already existed some theoretical
studies on the decay $\eta_{c2}\to J/\psi(\psi')+\gamma$ by
Sebastian and his
coworkers~\cite{Sebastian:1988kw,Sebastian:1994pu,Sebastian:1996cy}.
These authors worked in the traditional first-quantized
quantum-mechanical framework. Notably, some of serious
inconsistencies seem to exist amongst a sequence of their papers~\footnote{For example,
the partial width of $\eta_{c2}\to J/\psi+\gamma$ was originally
predicted to be $2.13$ keV~\cite{Sebastian:1988kw}. This prediction
later shifted to an abnormally large value, $62.6$ keV~\cite{Sebastian:1994pu}.
In their last publication~\cite{Sebastian:1996cy},
this prediction has shrunk to $0.699$ keV.}.
Therefore we feel it timely and obligatory to conduct an independent
investigation on these transition processes.

A modern effective-field-theory framework of dealing with the
single-photon transition in quarkonium has recently been put forward
in Ref.~\cite{Brambilla:2005zw}, by coupling the potential
nonrelativistic QCD (pNRQCD)~\cite{Pineda:1997bj} with the
electromagnetism. In \cite{Brambilla:2005zw}, the $M1$ transition
process with $\Delta l=0$, exemplified by $J/\psi\to \eta_c\gamma$,
has been systematically analyzed. In this work, we will utilize the
same formalism to tackle the $\Delta l=2$ magnetic transition
process $\eta_{c2}\to J/\psi(\psi')\gamma$. Our final results turn
out to significantly differ from those in \cite{Sebastian:1988kw,Sebastian:1994pu,Sebastian:1996cy}.

To account for the radiative transitions, it is convenient to
promote the gauge group of NRQCD to $SU(3)_{\rm c}\times
U(1)_{\em}$. One then matches this NRQCD action onto an even lower
energy effective field theory, pNRQCD, by further integrating out
the quantum fluctuations of virtuality of order $m^2v^2$. pNRQCD is
an ideal formalism to tackle radiative transitions, because the
active degrees of freedom, {\it i.e.}, the dynamical gluons and the
emitted photon, are both ultrasoft ($k^\mu\sim mv^2$) and can be
treated on an equal footing. Since they possess a wavelength much
longer than the typical quarkonium radius, the corresponding fields
need be multipole-expanded. By this way one can elegantly implement
the {\it multipole expansion}, the standard treatment of the
electromagnetic transitions in quantum mechanics textbooks.

Out starting point is the following pNRQCD Lagrangian density~\cite{Brambilla:2005zw}:
\bqa {\cal L}_{\rm pNRQCD} &=& \int d^3r \;  {\rm Tr} \, \Bigg\{
{\rm S}^\dagger \left( i\partial_0 + \frac{\bfnabla_R^2}{4 m}
+\frac{\bfnabla_r^2}{m}  - V_S^{(0)}(r) \right) {\rm S}\Bigg\} + {\cal
L}_{\rm light} + {\cal L}_{\gamma\, \rm pNRQCD} \,,
\label{pNRQCD:Lag}
\eqa
where $S$ represents the field for a composite system made of the
heavy quark and the heavy antiquark, depending on the center-of-mass
coordinate, $\bf R$, and the relative coordinate, $\bf r$. $S$ is a
$2 \times 2$ matrix in spinor space and a singlet under color and
$U(1)_{\rm em}$ transformations. The $V_S^{(0)}$ denotes the static
singlet $Q\overline{Q}$ potential. The first term in
(\ref{pNRQCD:Lag}), describing the dynamics of the $Q\overline{Q}$
pair dictated by the strong interaction, resembles very much the
traditional potential model. ${\cal L}_{\rm light}$ represents the
lagrangian for light degrees of freedom, {\it e.g.}, light quarks,
gluons as well as photons. The particularly relevant to this work is
the third term, ${\cal L}_{\gamma\, \rm pNRQCD}$, which depicts the
spin-dependent interaction between the $Q\overline{Q}$ pair and a
photon:
\bqa
{\cal L}_{\gamma\, \rm pNRQCD} &=&  \int d^3 r \; {\rm Tr} \,
\Bigg\{ \; e e_Q {\rm S}^\dagger {\bf r}\cdot {\bf E}^{\em} {\rm S}
+ {c_F^\em  e e_Q\over 2 m} \; \left\{{\rm S}^\dagger , \bfsigma
\cdot {\bf B}^{\em}\right\} {\rm S}
\nn \\
&& + \frac{c_F^\em  e e_Q }{16 m} \; \left\{ {\rm S}^\dagger ,
\bfsigma \cdot  ({\bf r}\cdot \bfnabla_R)^2{\bf B}^{\em}\right\}
{\rm S} + {e e_Q \over 8 m^2} \; r V_S^{(0)\prime} \; \left\{{\rm
S}^\dagger , \bfsigma\cdot \hat{\bf r} \times  \left(
  \hat{\bf r}\times {\bf B}^{\em} \right) \right\} {\rm S}
\nn \\
& & - {  c_S^\em e e_Q \over 16 m^2} \; \left[{\rm S}^\dagger,
\bfsigma \cdot \left[-i\bfnabla_R \times, {\bf E}^\em \right]\right]
{\rm S} - {c_S^\em e e_Q  \over 16 m^2} \; \left[ {\rm S}^\dagger,
\bfsigma \cdot \left[-i\bfnabla_r \times, ({\bf r} \cdot \bfnabla_R)
{\bf E}^\em \right]\right] {\rm S}
\nn \\
&&  + {c^\em_{W12} e e_Q \over 4 m^3} \; \left\{ {\rm S}^\dagger ,
\bfsigma \cdot  {\bf
  B}^{\em} \right\} \bfnabla_r^2 {\rm S}
+ {c^\em_{p^\prime p} e e_Q  \over 4 m^3}
\; \left\{ {\rm S}^\dagger , \bfsigma^i \, {\bf
  B}^{\em\,j} \right\} \bfnabla_r^i\bfnabla_r^j {\rm S} \Bigg\}\,,
\label{gammapNRQCD:Lag}
\eqa
where the trace is over the spin indices. The ${\bf E}^\em$ and
${\bf B}^\em$ signify the electric and magnetic field strengths,
which depend only on $\bf R$. The ubiquitous occurrence of the Pauli
matrix $\bfsigma$ (the first operator, responsible for the $E1$
transition, is an exception), signals that the concerned radiative
transition is of the spin-dependent type, {\it i.e.}, with quark
spin flipped. The appearance of $\bf r$ is a consequence of
multipole-expanding the electricmagnetic field, while the $\hat{\bf
r}\equiv {\bf r}/r$ represents the unit radial vector. Notice that
(\ref{gammapNRQCD:Lag}) is manifestly gauge invariant.

The coefficients $c_F^\em$, $c_S^\em$, $c^\em_{W12}$, and
$c^\em_{p^\prime p}$ in (\ref{gammapNRQCD:Lag}) are the matching
coefficients that were directly inherited from the NRQCD with
enlarged gauge group. They satisfy some exact relations dictated by
reparametrization (or Poincar\'e) invariance \cite{Manohar:1997qy}:
\bqa
c_S^\em &=& 2 c_F^\em -1, \qquad  c_{W12}^\em \equiv  c_{W1}^\em
-c_{W2}^\em =1,\qquad c_{p^\prime p}^\em = c_F^\em -1.
\label{rpi:matching:coeff}
\eqa
All these matching coefficients are known at the one loop
level~\cite{Manohar:1997qy}. In particular, we have
\bqa
 c_F^\em \equiv 1+ \kappa_Q^\em &=& 1+{4\over 3}
\frac{\als}{2\pi} + {\cal O}(\als^2)\,,
\label{CF:kappa:Q}
\eqa
where $\kappa_Q^\em$ is usually dubbed the {\it anomalous magnetic
moment} of the heavy quark. The one-loop perturbative contribution
to $\kappa_Q^\em$ in (\ref{CF:kappa:Q}) seems insignificant, less
than 10\% for charm. However, it is often conjectured that the
magnetic moment of the bound quark may receive a large contribution
due to some nonperturbative mechanism. For simplicity we will
suppress such a possibility (see \cite{Brambilla:2005zw} for a
discussion on whether such a contribution can naturally emerge from
the first principle of QCD). Rather we will content ourselves with
taking $\kappa_Q$ as its short-distance value.

The $\bfsigma\cdot \hat{\bf r} \times (\hat{\bf r}\times {\bf
B}^{\em})$ term in (\ref{gammapNRQCD:Lag}) is also worth some
remarks. As discussed in \cite{Brambilla:2005zw}, unlike the
remaining operators listed in (\ref{gammapNRQCD:Lag}), this
spin-dependent transition operator receives some nontrivial
corrections when descending from NRQCD to pNRQCD. This operator is
intimately related to the quark spin-orbit potential, whose
coefficient is also protected by the reparametrization (or
Poincar\'e) invariance (Gromes relation)~\cite{Gromes:1984ma}.

Prior to giving concrete expression for $\eta_{c2}\to
J/\psi(\psi')\gamma$, we note that there have been some
phenomenological inclinations that $\psi'$ may not be a pure $S$-wave
vector charmonium. Rather it may have some admixture of ${}^3D_1$
component~\cite{Rosner:2001nm}:
\bqa
|\psi'\rangle &=& \cos\phi |2 ^3S_1 \rangle -\sin \phi |1
^3D_1\rangle, \label{S:D:mixing}
\eqa
usually with $\phi$ taken to be around $12^\circ$.

To compute the radiative transition process in pNRQCD, it is
necessary to know the corresponding vacuum-to-quarkonium pNRQCD
matrix elements. To the purpose of this work, we need know
\begin{subequations}
\bqa
\langle 0 |S({\bf R},{\bf r})|n^3S_1({\bf P},\lambda) \rangle  &=&
\frac{1}{\sqrt{4\pi}} \, R_{n0}(r)  \, \frac{\bfsigma \cdot {\bf
e}_{n^3S_1}(\lambda)}{\sqrt{2}}\,e^{i{\bf P}\cdot {\bf R}},
\label{wavefun:jpsi}
\\
\langle 0 |S({\bf R},{\bf r})|n^1P_1({\bf P},\lambda) \rangle &=&
\sqrt{\frac{3}{4\pi}} \, R_{n1}(r)  \, \frac{{\bf
e}_{n^1P_1}(\lambda)\cdot{\hat{\bf r}} }{\sqrt{2}}\,e^{i{\bf P}\cdot
{\bf R}}, \label{wavefun:hc}
\\
\langle 0 |S({\bf R},{\bf r})|n^1D_2({\bf P},\lambda) \rangle  &=&
\sqrt{\frac{15}{8\pi}} \, R_{n2}(r)  \, {\hat{r}^i  \,
h^{ij}_{n^1D_2}(\lambda) \, \hat{r}^j \over \sqrt{2}}\,e^{i{\bf
P}\cdot {\bf R}}, \label{wavefun:etac2}
\\
\langle 0 |S({\bf R},{\bf r})|n^3D_1({\bf P},\lambda) \rangle &=& {3
\over \sqrt{8\pi}} \, R_{n2}(r) \, { {\bf e}(\lambda)\cdot \hat{\bf
r} \, \bfsigma \cdot \hat{\bf r}-{1\over 3}\,{\bf e}(\lambda)\cdot
\bfsigma \over \sqrt{2}}\,e^{i{\bf P}\cdot {\bf R}}.
\label{wavefun:3D1}
\eqa
\label{vac:to:quark:pNRQCD:me}
\end{subequations}
The first two entries have been constructed in
\cite{Brambilla:2005zw}, and the last two are new.
Here $R_{nl}(r)$ denotes the radial Schr\"{o}dinger wave function of
a quarkonium. The symbol ${\bf e}(\lambda)$ represents the
polarization vector of any $J=1$ quarkonium state, 
satisfying the
orthogonality condition ${\bf e}^*(\lambda)\cdot {\bf
e}(\lambda^\prime) = \delta_{\lambda\lambda'}$, whereas the symbol
$h^{ij}(\lambda)$ denotes the polarization tensor of the $n^1D_2$
state, which is symmetric and traceless, obeying the orthogonality
condition ${\rm tr}\big[h^{*}(\lambda) h(\lambda^\prime)\big] =
\delta_{\lambda\lambda'}$. The overall normalization factors for
each entity are chosen such that all the quarkonium states are
nonrelativistically normalized.

With the knowledge of (\ref{gammapNRQCD:Lag}) and
(\ref{vac:to:quark:pNRQCD:me}), it is then a straightforward
exercise to calculate the transition amplitude for ${}^1D_2 \to
{}^3S_1+\gamma$ and ${}^1D_2 \to {}^3D_1+\gamma$. The latter is a
regular $M1$ transition between the $D$-wave spin singlet and
triplet, which is included here because it can contribute to
$\eta_{c2}\to \psi'+\gamma$ through the mixing mechanism. After completing
the angular integration, the
desired results are
\begin{subequations}
\bqa
{\mathcal M}[ {}^1D_2 \to {}^3S_1+\gamma] &=& {e e_Q \over 2 m_Q}
\sqrt{2\over 15}\bigg\{ { {\bf e}^* \cdot {\bf k} \times
\bm{\varepsilon}^*_\gamma\, k^i h^{ij}k^j \over |{\bf k}|^2} c_F^\em
J_1 + k^i h^{ij}\left(  {\bf e}^* \times \bm{\varepsilon}^*_\gamma
\right)^j (c_S^\em-1) J_2
\nn\\
&+& e^{*i} h^{ij} \left({\bf k} \times
\bm{\varepsilon}^*_\gamma\right)^j \left( J_2+J_4 - c_{p'p}^\em J_3
\right) \bigg\},
\label{D:S:tran:ampl}
\\
{\mathcal M}[ {}^1D_2 \to {}^3D_1+\gamma] &=& {e e_Q \over 2 m_Q} {6
\over \sqrt{15}}\, e^{*i} h^{ij} \left({\bf k} \times
\bm{\varepsilon}^*_\gamma\right)^j \,c_F^\em J_0,
\label{D:D:M1:tran:ampl}
\eqa
\label{etac2:jpsi:trans:ampl}
\end{subequations}
where ${\bf k}$ and $\bm{\varepsilon}^*_\gamma$ represent the three-momentum and polarization 
vector of the emitted photon, respectively, and ${\bf e}^*$ stands for that of the 
outgoing $J=1$ charmonium.
The involved {\it dimensionless} overlap integrals are given by
\begin{subequations}
\bqa
J_0&=&   \int^\infty_0 \!\!\! dr\: R_{^3D_1}(r) R_{^1D_2}(r)\,r^2,
\\
J_1&=& -{|{\bf k}|^2\over 4}  \int^\infty_0 \!\!\! dr\:
R_{n^3S_1}(r) R_{^1D_2}(r)\,r^4,
\\
J_2&=& {|{\bf k}|\over 2 m_Q} \int^\infty_0 \!\!\! dr\: R^\prime_{n^3S_1}(r) R_{^1D_2}(r)\,r^3,
\\
J_3 &=& - {1 \over m_Q^2} \int^\infty_0  \!\!\! dr\: \left(R^{\prime\prime}_{n^3S_1}(r)
-{R^\prime_{n^3S_1}(r)\over r}\right)R_{^1D_2}(r)\,r^2,
\\
J_4 &=&  {1 \over 2 m_Q}  \int^\infty_0  \!\!\! dr\:
R_{n^3S_1}(r) V_S^{(0) \prime}(r)
R_{^1D_2}(r)\,r^3.
\eqa
\label{overlap:integrals:J0:J4}
\end{subequations}
To facilitate the comparison with the expressions in
Ref.~\cite{Sebastian:1988kw,Sebastian:1994pu,Sebastian:1996cy}, we
have introduced the same $J_i$ ($i=0,\ldots,4$) as theirs, except a
different normalization factor for $J_1$ is adopted.

Some remarks on deriving the amplitude (\ref{D:S:tran:ampl}) are in
order. Unlike in the case of an ordinary $M1$ transition process
with $\Delta l=0$, not all of the operators in
(\ref{gammapNRQCD:Lag}) can make a nonvanishing contribution for
such a $\Delta l=2$ transition process. For example, the leading
magnetic-dipole operator, the $\bfsigma \cdot {\bf B}^{\em}$ term,
and the $\bfsigma \cdot \left[-i\bfnabla_R \times, {\bf E}^\em
\right]$ term, as well as the $\bfsigma \cdot {\bf B}^{\em}
\bfnabla_r^2$ term, simply cannot connect a $D$-wave state to the
$S$-wave state owing to their spherically-symmetrical nature. It
turns out that only four terms in (\ref{gammapNRQCD:Lag}) can
directly make a nonzero contribution.

Equation (\ref{D:S:tran:ampl}) has also embedded an interesting
piece of contribution, the so-called {\it final-state recoil}
correction~\cite{Grotch:1982bi}. As a manifestation of the
relativistic effect to a moving ${}^3S_1$ state, its wave function
can develop a nonvanishing overlap with a spin-singlet $P$-wave
component, therefore the transition ${}^1D_2\to {}^3S_1$ can be
effective realized through a $E1$ transition from the parent to this
small ${}^1P_1$ component. Some subtle gauge-invariance issue
related to this Lorentz boost effect has been
clarified~\cite{Brambilla:2005zw}.
Although the recoil correction and
the spin-orbit-potential-related term are not separately invariant
under the $U(1)_\em$ transformation, their sum is.
We have explicitly verified that, for the reaction ${}^1D_2 \to {}^3S_1+\gamma$,
the sum of the recoil correction and the correction induced by
the spin-orbit-potential-related operator,
is indeed independent of the redefinition of the pNRQCD field
$S$ by implementing a $U(1)_\em$ gauge link.

For the decay ${}^1D_2 \to {}^3D_1+\gamma$, which is an ordinary
allowed $M1$ transition, we only consider the leading contribution
and not include the relativistic correction in
(\ref{D:D:M1:tran:ampl}).

Squaring the amplitudes in (\ref{etac2:jpsi:trans:ampl}), and
summing over all possible polarizations, it is easy to obtain the
spin-averaged partial width $\Gamma[\eta_{c2}\to
J/\psi(\psi')+\gamma]$. Nevertheless, (\ref{etac2:jpsi:trans:ampl})
encodes much richer polarization information. In this work, we
would like to proceed to extract the helicity
amplitude~\cite{Jacob:1959at,Martin:1970} and the multipole
amplitude~\cite{Karl:1975jp,Karl:1980wm} from this equation.
These two types of amplitudes can in principle be extracted 
experimentally. It is worth noting that, 
CLEO-c experiment has recently extracted the higher-order 
multipole amplitudes associated the cascade decay process 
$\psi^\prime\to \chi_{c1,2}\gamma\to J/\psi+\gamma\gamma$,
by performing a maximum likelihood fit of the joint angular 
distributions of the two photons~\cite{:2009pn}.

For the decay $\eta_{c2}(\nu)\to \psi(\mu)+\gamma(\lambda)$, we
signify the projection of the angular momentum of $\eta_{c2}$ along
the moving direction of the $\gamma$ by $\nu$, and denote the
helicities of $\psi$ and $\gamma$ by $\mu$ and $\lambda$. With this
specific choice of the quantization axis, we have $\nu=\lambda-\mu$.
There are in total three independent helicity amplitudes
$A_{\mu,\lambda}$, and the other three can be related by parity
invariance. We introduce the shorthand $A_\nu$ for
$A_{\mu,\lambda}$ with $\lambda$ fixed to be $+1$. Substituting the
explicit representation of polarization tensors in
(\ref{etac2:jpsi:trans:ampl}), it is easy to obtain:
\begin{subequations}
\bqa
A_0 &\equiv& A_{1,1}= -A_{-1,-1}= {2 c_F^\em J_1-(2
c_S^\em-1)J_2+c_{p^\prime p}^\em J_3-J_4 +3 \sqrt{2} c_F^\em
   J_0 \sin\phi \over \sqrt{6}},
\\
A_1 &\equiv& A_{0,1} = -A_{0,-1}= {-c_S^\em J_2 + c_{p^\prime p}^\em
J_3- J_4 + 3 \sqrt{2} c_F^\em J_0 \sin\phi \over \sqrt{2}},
\\
A_2 & \equiv & A_{-1,1}= -A_{1,-1} = -J_2+ c_{p^\prime p}^\em
   J_3 - J_4 + 3 \sqrt{2} c_F^\em J_0\sin\phi,
\eqa
\label{etac2:jpsi:gamma:helicty:ampl}
\end{subequations}
where $\phi$ is the ${}^3S_1\!-\!{}^3D_1$ mixing angle, which equals $0$ for $J/\psi$,
and may be put $12^\circ$ for $\psi^\prime$ on phenomenological ground. 
In deriving the helicity amplitudes for
$\psi^\prime+\gamma$, we have approximated $\cos\phi=0.978\approx 1$ for simplicity.

From (\ref{etac2:jpsi:gamma:helicty:ampl}), it is straightforward to
obtain the spin-averaged partial width:
\bqa
\Gamma[\eta_{c2}\to J/\psi(\psi')+\gamma] &=& \frac{2\alpha e_Q^2
|{\bf k}|^3}{75m_c^2}\bigg(|A_0|^2+|A_1|^2+|A_2|^2\bigg).
\label{partial:width:etac2:jpsi:gamma:1}
\eqa

According to \cite{Karl:1975jp,Karl:1980wm}, the helicity amplitudes
$A_\nu$ ($\nu=0,1,2$) are connected to the multipole amplitudes
$a_{J_{\gamma}}$  through the following
orthogonal transformation~\footnote{Note that this transformation matrix is identical
to the corresponding one in the $\chi_{c2}\to J/\psi+\gamma$ process~\cite{:2009pn}. 
But for such a $E1$-dominant process, $a_1, a_2, a_3$ 
should be identified with the electric dipole ($E1$), 
magnetic quadrupole ($M2$), and electric octupole ($E3$) amplitudes, 
respectively.}:
\begin{equation}
A_\nu=\sum_{J_\gamma} \sqrt{2 J_\gamma + 1 \over 2 J_{\eta_{c2}}+1}
\, a_{J_\gamma} \,\langle
J_\gamma,1;1,\nu-1|J_{\eta_{c2}},\nu\rangle,
\label{A:a:transform}
\end{equation}
where the Condon-Shortley notation for the Clebsch-Gordan
coefficients, $\langle j_1,m_1;j_2,m_2|J,M\rangle$, has been
adopted~\cite{Amsler:2008zzb}. $J_{\eta_{c2}}=2$ is the spin of the
$\eta_{c2}$ meson. $J_\gamma$ represents the angular
momentum carried by the photon, obeying $1 \leq J_\gamma \leq J_{\eta_{c2}}+1$,
and $a_1$, $a_2$ and $a_3$ correspond to the $M1$, $E2$, and $M3$
multipole amplitudes, respectively.

The inverse transformation of (\ref{A:a:transform}) can be readily
obtained:
\begin{equation}
\left(
\begin{array}{c}
a_1\\
a_2\\
a_3
\end{array}
\right)
=
\left(
\begin{array}{ccc}
 \frac{1}{\sqrt{10}} & \sqrt{\frac{3}{10}} & \sqrt{\frac{3}{5}}
   \\
 \frac{1}{\sqrt{2}} & \frac{1}{\sqrt{6}} & -\frac{1}{\sqrt{3}}
   \\
 \sqrt{\frac{2}{5}} & -2 \sqrt{\frac{2}{15}} &
   \frac{1}{\sqrt{15}}
\end{array}
\right)
\left(
\begin{array}{c}
A_0\\
A_1\\
A_2
\end{array}
\right).
\end{equation}

From (\ref{etac2:jpsi:gamma:helicty:ampl}), we can readily deduce
these three multipole amplitudes:
\begin{subequations}
\bqa
a_{1}&=& {2 c_F^\em J_1-5 (1+c_S^\em) J_2 +10 c_{p^\prime p}^\em
J_3-10
   J_4 + 30
   \sqrt{2} c_F^\em J_0 \sin\phi \over 2 \sqrt{15}},
\\
a_{2}&=& {2 c_F^\em J_1-3 (c_S^\em-1) J_2 \over 2 \sqrt{3}},
\\
a_3&=& {2 c_F^\em J_1 \over \sqrt{15}}.
\eqa
\label{etac2:jpsi:gamma:multipole:ampl}
\end{subequations}
As expected, the $S\!-\!D$-wave mixing effect can only enter into the
$M1$ contribution. Note the $M3$ amplitude solely receives the
contribution from the $\bfsigma \cdot ({\bf r}\cdot
\bfnabla_R)^2{\bf B}^{\em}$ operator in (\ref{gammapNRQCD:Lag}),
which comes from multipole-expanding the leading magnetic transition
operator to the second order in $\bf r$.

We are now in a position to compare our expressions for the three
multipole amplitudes (\ref{etac2:jpsi:gamma:multipole:ampl}) with
those reported in \cite{Sebastian:1994pu,Sebastian:1996cy}. Somewhat
surprisingly, it seems that our expressions
disagree with theirs for each multipole,
except for the $J_0$ and $J_1$ pieces.
The authors of \cite{Sebastian:1994pu,Sebastian:1996cy} have not included the
anomalous magnetic moment of the $c$ quark, and in their formulas,
all the overlap integrals, $J_1$ through $J_4$, contribute to the
$M1$ and $E2$ amplitudes. Taking $\kappa_Q=0$ in
(\ref{etac2:jpsi:gamma:multipole:ampl}), we find that $J_3$ would
disappear from the $M1$ amplitude, and the $E2$ amplitude will only
depend on $J_1$, which are in diametric contradiction to the
equations in \cite{Sebastian:1994pu,Sebastian:1996cy}.

We can express the spin-averaged transition width for $\eta_{c2}\to
J/\psi(\psi')+\gamma$ in terms of three multipole amplitudes:
\bqa
\Gamma[\eta_{c2}\to J/\psi(\psi')+\gamma] &=& {2 \alpha e_Q^2 |{\bf
k}|^3 \over 75m_c^2 }\bigg(|a_1|^2+|a_2|^2+|a_3|^2\bigg),
\eqa
which is equivalent to (\ref{partial:width:etac2:jpsi:gamma:1}),
since the transformation (\ref{A:a:transform}) preserves the
norm.

To make concrete predictions for the transitions $\eta_{c2}\to
J/\psi(\psi')+\gamma$, we need know the radial wave functions of
each involved charmonium from some phenomenological potential
models. To ensure that our conclusion not to rest heavily upon one
specific model, we choose to study with five different potential
models. All of them only differ in the way of parameterizing the
static potential $V_S^{(0)}$. Specifically, the potentials we choose
are Cornell type~\cite{Eichten:1978tg}, the Buchm\"{u}ller-Tye (BT)
type~\cite{Buchmuller:1980su}, NR potential by Barnes {\it et
al.}~\cite{Barnes:2005pb}, the screened confinement potential by Li
and Chao~\cite{Li:2009zu}, and the potential proposed by
Fulcher~\cite{Fulcher:1991dm}. We solve the Schr\"{o}dinger equation
with these potentials numerically, with the input parameters taken
from the aforementioned papers.

In Table~\ref{table1}, we tabulate the predictions from the various
models for the overlap integrals $J_i$ ($i=1,\cdots,4$). When
evaluating these integrals defined in
(\ref{overlap:integrals:J0:J4}), we have used $m_c=1.5$ GeV, and
determined the photon momentum by physical kinematics, {\it i.e.},
we assume $\eta_{c2}$ with a mass of 3872 MeV and use the physical
masses of $J/\psi$ and $\psi'$ as input. Thus we obtain $|{\bf
k}|=698$ MeV~\footnote{One may argue that,
the $700$ MeV photon in $\eta_{c2}\to J/\psi\gamma$
is a little bit too energetic, and it is perhaps
more appropriate to count $|{\bf k}|\sim mv$ in this case.
This may cast some doubt on the validity of multipole expansion,
the underlying basis of this work.
It might be illuminating to apply the {\it hard-scattering mechanism},
 recently developed in \cite{Jia:2009yg}, to reinvestigate this channel.}
for $\eta_{c2}\to J/\psi\gamma$ and 181 MeV for
$\eta_{c2}\to \psi'\gamma$. From Table~\ref{table1}, it is
reassuring that these overlap integrals are not very sensitive to
the different models. On the other hand, the overlap integral $J_0$
can be safely put to be unity, since to the intended accuracy, both
the $\eta_{c2}(1D)$ and the $\psi(1D)$ have degenerate radial wave
function.

In Table~\ref{table2}, we give the explicit predictions for the
decay $\eta_{c2}\to J/\psi(\psi')+\gamma$, such as the partial
width, the helicity amplitudes and the (normalized) multipole
amplitudes for each decay channel within various potential models.
We have taken $\alpha=1/137$ and the electric charge of the charm quark,
$e_c=2/3$. We have used the one-loop perturbative value for the anomalous
magnetic momentum of the $c$, as indicated in (\ref{CF:kappa:Q}). If
 $\alpha_s(m_c)=0.35$ is used, we then get $\kappa_c=0.074$.
Detailed numerical checks reveal that all the predictions in
Table~\ref{table2} only change modestly if we set $\kappa_c=0$.

From Table~\ref{table2}, one finds that the predicted partial width for
$\eta_{c2}\to J/\psi\gamma$ from various potential models ranges between $3.11$ and $4.78$ keV,
that for $\eta_{c2}\to \psi^\prime \gamma$ without considering ${}^3S_1\!-\!{}^3D_1$ mixing
ranges from $0.017$ to $0.029$ keV, and that for $\eta_{c2}\to \psi^\prime \gamma$ incorporating
the mixing effect ($\phi=12^\circ$) ranges from $0.49$ to $0.56$ keV.
This seems to indicate that different potential models give rise to
reasonably consistent predictions for each channel. 

\begin{table}[!hbp]
\centering \caption{{\label{table1}} The overlap integrals $J_i$ for
the electromagnetic transition $\eta_{c2}\to J/\psi (\psi')\gamma$
in various potential models. Since $J_0=1$, we have not listed its
value.}
\begin{tabular}{|c|cc|cc|cc|cc|cc}
\hline \backslashbox{$J_i$}{\\Potential\\
Models}&\multicolumn{2}{c|}{Cornell~\cite{Eichten:1978tg}
}&\multicolumn{2}{c|}{Screened~\cite{Li:2009zu}
}&\multicolumn{2}{c|}{NR~\cite{Barnes:2005pb}
}&\multicolumn{2}{c|}{BT~\cite{Buchmuller:1980su}
}&\multicolumn{2}{c}{Fulcher~\cite{Fulcher:1991dm}}\\
\hline &$J/\psi$ & $\psi^\prime$ & $J/\psi$ & $\psi^\prime$&$J/\psi$
&
$\psi^\prime$& $J/\psi$ & $\psi^\prime$&$J/\psi$ & $\psi^\prime$\\
\hline
$J_1$ &-0.600 & 0.123 & -0.710 &0.180 & -0.728 & 0.161 & -0.757 & 0.153 & -0.763 &0.147\\
\hline
$J_2$&-0.376&0.051&-0.347&0.063&-0.365&0.056&-0.383&0.048&-0.390&0.044\\
\hline
$J_3$&-0.304&-0.256&-0.227&-0.160&-0.242&-0.191&-0.245&-0.212&-0.249&-0.225\\
\hline
$J_4$&0.136&-0.243&0.121&-0.218&0.128&-0.231&0.144&-0.244&0.173&-0.291\\
\hline
\end{tabular}
\end{table}

\begin{table}[!hbp]
\centering \caption{{\label{table2}}  The predictions of
$\eta_{c2}\to J/\psi(\psi')+\gamma$ from various potential models.
The mixing angle $\phi$ has been taken for both $0^\circ$ and $12^\circ$
for $\psi'$. We have taken $\alpha=1/137$, and $\kappa_c=0.074$ by
using $\alpha_s(m_c)=0.35$. In addition to the spin-averaged partial width, the
helicity amplitudes and the (normalized) multipole amplitudes for
each decay channel have also been given.}
\begin{tabular}{c|c|cccccccccc}
 \hline
Potential Models &&$\phi$ &$A_0$& $A_1$ & $A_2$ & $a_1$ & $a_2$
&$a_3$
&$|a_2/a_1|$ & $|a_3/a_1|$ & Width (keV)\\
 \cline{1-12}
 \multicolumn{1}{c|}{\multirow{3}*{Cornell}} &
$J/\psi$&-- &-0.39&0.19&0.22&0.31&-0.66&-0.68&2.15 & 2.21 &   3.11 \\
\cline{2-12}
 \multicolumn{1}{c|}{}
&\multicolumn{1}{c|}{\multirow{2}*{$\psi^\prime$}}&
$0^\circ$&0.17&0.12&0.17&0.93&0.26&0.25&0.28&0.27&0.017\\\cline{3-12}
 \multicolumn{1}{c|}{}      &\multicolumn{1}{c|}{} &
 $12^\circ$&0.56&0.79&1.12&0.99&0.05&0.05&0.05&0.05&0.50
 \\
\hline
 \multicolumn{1}{c|}{\multirow{3}*{Screened}} &
$J/\psi$&--&-0.50&0.18&0.21&0.19&-0.70&-0.70&3.72&3.70&4.22
\\\cline{2-12}
  \multicolumn{1}{c|}{}      &
\multicolumn{1}{c|}{\multirow{2}*{$\psi^\prime$}}&
 $0^\circ$&0.21&0.09&0.14&0.85&0.38&0.37&0.45&0.44&0.017\\\cline{3-12}
 \multicolumn{1}{c|}{}       & \multicolumn{1}{c|}{} &
  $12^\circ$&0.60&0.76&1.09&0.99&0.07&0.07&0.07&0.07&0.49
 \\
        \hline
 \multicolumn{1}{c|}{\multirow{3}*{NR}} & $J/\psi$
&--&-0.50&0.19&0.22&0.20&-0.69&-0.69&3.49&3.49&4.45\\\cline{2-12}
\multicolumn{1}{c|}{}
&\multicolumn{1}{c|}{\multirow{2}*{$\psi^\prime$}}& $0^\circ$&
0.20&0.11&0.16&0.89&0.33&0.32&0.38&0.36&0.018\\\cline{3-12}
\multicolumn{1}{c|}{}         & \multicolumn{1}{c|}{} & $12^\circ$&
0.59&0.78&1.11&0.99&0.06&0.06&0.06&0.06&0.50
\\
        \hline
 \multicolumn{1}{c|}{\multirow{3}*{BT}} & $J/\psi$
&--&-0.53&0.20&0.22&0.18&-0.70&-0.69&3.76&3.76& 4.78\\\cline{2-12}
  \multicolumn{1}{c|}{}
&\multicolumn{1}{c|}{\multirow{2}*{$\psi^\prime$}}&
 $0^\circ$ &0.20 &0.12&0.18&0.91&0.30&0.29&0.33&0.31&0.020 \\\cline{3-12}
  \multicolumn{1}{c|}{}       & \multicolumn{1}{c|}{} &
  $12^\circ$&0.59 &0.79&1.13&0.99&0.06&0.06&0.06&0.06&0.51
  \\
 \hline
 \multicolumn{1}{c|}{\multirow{3}*{Fulcher}} & $J/\psi$
&--&-0.54&0.18&0.20&0.14&-0.70&-0.70&5.16&5.16&4.77\\\cline{2-12}
 \multicolumn{1}{c|}{}       &
\multicolumn{1}{c|}{\multirow{2}*{$\psi^\prime$}}&
  $0^\circ$&0.22&0.16&0.23&0.94&0.24&0.23&0.26&0.24&0.029  \\\cline{3-12}
   \multicolumn{1}{c|}{}      & \multicolumn{1}{c|}{} &
    $12^\circ$&0.60&0.83&1.18&0.99&0.05&0.05&0.05&0.05&0.56
   \\
 \hline
\end{tabular}
\end{table}

In contrast to the minor role played by retaining $\kappa_c$, the effect of ${}^3S_1\!-\!{}^3D_1$ mixing is enormous for
the decay $\eta_{c2}\to \psi'\gamma$, as can be clearly seen from
Table~\ref{table2}. The predicted partial widths with
$\phi=12^\circ$ are about $25$ times larger than those without
including the mixing for $\psi'$. This can be attributed to the fact
that the transition $^1D_2 \to {}^3D_1 \gamma$ is an allowed $M1$
transition, with the overlap integral $J_0=1$. However for the
transition $^1D_2 \to  2{}^3S_1+\gamma$, due to the existence of a
node in the $2S$ radial wave function, the corresponding overlap
integrals are generally small. Therefore, even with a relatively
small $\phi$ angle, the $S\!-\!D$-wave mixing can already play a very
important role.

In Table~\ref{table2}, we also list the magnitudes of the $M1$,
$E2$, $M3$ amplitudes for each transition process. For clarity, we have
renormalized $a_i$ ($i=1,2,3$) in
(\ref{etac2:jpsi:gamma:multipole:ampl}) such that the new multipole
amplitudes satisfy $|a_1|^2+|a_2|^2+|a_3|^2=1$.
It is a very interesting observation that $E2$ and $M3$ amplitudes are
almost identical in {\it all} the decay channels.
For $\eta_{c2}\to J/\psi\gamma$, these two higher-order multipoles are
even more important in magnitude than the $M1$ amplitude!
This pattern no longer holds true for the $\eta_{c2}\to \psi'\gamma$ channel.
Nevertheless, it is also interesting to
note that, for $\eta_{c2}\to \psi'\gamma$, the ratios
$|a_2/a_1|$ and $|a_3/a_1|$ decrease significantly after the mixing
effect is included. This can also be understood by the fact that,
since the allowed $M1$ transition $^1D_2 \to {}^3D_1 \gamma$ makes
more pronounced contribution than $^1D_2 \to {}^3S_1 \gamma$, 
including the mixing effect thus can greatly amplify the importance of 
the $M1$ amplitude relative to other two
multipoles.

Surveying Table~\ref{table2}, one may be able to place the
following saturation bound for the ratio of the two branching fractions:
\begin{subequations}
\bqa
&& { {\cal B}[\eta_{c2}\to \psi'+ \gamma] \over {\cal
B}[\eta_{c2}\to J/\psi+\gamma] } \leq 0.16, 
\quad\quad\quad\phi=12^\circ
\\
&& {{\cal B}[\eta_{c2}\to \psi'+ \gamma] \over {\cal B}[\eta_{c2}\to
J/\psi+\gamma]} \leq 6.1\times 10^{-3},\quad\quad\quad   \phi=0^\circ
\label{ratio:braching fractions}
\eqa
\end{subequations}
where the maximum ratio in first equation comes from the prediction in Cornell potential,
and that in second equation is obtained from Fulcher's potential model.
These results seem to seriously conflict with the corresponding
\textsc{BaBar} measurements, (\ref{ratio:two:branch:ratio:Babar}), no
matter the mixing effect for $\psi'$ is taken into account or not.
This seems to be a strong evidence to disfavor the $\eta_{c2}$
assignment for the $X(3872)$. At this stage, these upper limits seem
still compatible, but orders of magnitude less, with the preliminary
\textsc{Belle} upper bound on this ratio,
(\ref{ratio:two:branch:ratio:Belle}).

It is certainly interesting to examine whether the absolute
branching fractions of $\eta_{c2}\to J/\psi(\psi')\gamma$ are
compatible with the $B$ factory measurements or not. To realize this
goal, one first needs know the full width of the $X(3872)$.
\textsc{Belle} has set an upper limit for the total width of
$X(3872)$: $\Gamma_X <2.3$ MeV at 90\% confidence
level~\cite{Choi:2003ue}. The 2008 PDG compilation has estimated the
total width of the $X(3872)$ to be $\Gamma_X =3.0^{+2.1}_{-1.7}$
MeV~\cite{Amsler:2008zzb}.

We hope to establish an upper bound for the predicted branching
ratios of $\eta_{c2}\to J/\psi(\psi')\gamma$. To this purpose, we
have taken a conservative attitude, by assuming $\Gamma_{X}=1.3$ 
MeV, the lower end of the PDG estimate. Such a value by itself is
consistent with the full width of the $\eta_{c2}$. From
Table~\ref{table2}, one can see that, among all the potential models
analyzed in this work, the BT potential model
gives the largest prediction to the partial width of $\eta_{c2}\to
J/\psi \gamma$, $4.78$ keV. We thus estimate
\bqa
&& {\cal B}[\eta_{c2}\to J/\psi+ \gamma] \leq 3.7\times 10^{-3}.
\label{br:etac2:jpsi:photon:up:limit}
\eqa
It is interesting to contrast this theoretical upper bound
with those inequalities inferred from the two $B$ factory experiments,
(\ref{inequality:X:jpsi:psi2S:gamma:exp}). Equation (\ref{br:etac2:jpsi:photon:up:limit}) 
seems to mildly conflict with the experimental lower bound from either \textsc{BaBar} or \textsc{Belle}. 
Due to large theoretical uncertainties in $\Gamma[\eta_{c2}\to J/\psi+ \gamma]$ and $\Gamma_X$, 
we are inclined not to make strong claim just based on the analysis for this channel.

The decay $\eta_{c2}\to \psi'\gamma$ poses a much more stringent
challenge to the data. Among all the potential models, Fulcher's
potential model yields the largest prediction to the partial width of
$\eta_{c2}\to \psi' \gamma$, with or without including the mixing
effect. Taking the corresponding partial widths from
Table~\ref{table2}, we may place the following inequalities:
\begin{subequations}
\bqa
&& {\cal B}[\eta_{c2}\to \psi^\prime+ \gamma] \leq 4.3\times
10^{-4},\quad\quad\quad\phi=12^\circ
\\
&& {\cal B}[\eta_{c2}\to \psi^\prime+ \gamma] \leq 2.2\times
10^{-5}.\quad\quad\quad\phi=0^\circ
\eqa
\end{subequations}
Comparing these theoretical upper bounds with the lower bound
imposed by the \textsc{BaBar} measurement, $1.9 \times 10^{-2}$,
which is given in (\ref{inequality:X:jpsi:psi2S:gamma:exp}), one
observes the glaring discrepancies, no matter the mixing effect for
$\psi'$ is included or not. If \textsc{BaBar} result can be trusted,
this may also be viewed as a strong evidence against assigning the
$X(3872)$ as the $\eta_{c2}$.

\begin{table}[!hbp]
\centering  \caption{{\label{table3}} 
The predictions of the overlap integral $\langle r\rangle$  and the partial width 
for the electric transition $\eta_{c2} \to h_c+\gamma$ in various potential models.
The mass of $h_c$ is taken to be $3525$ MeV~\cite{Amsler:2008zzb}.}
\begin{tabular}{c|cc|cc|cc|cc|cc}
\hline \backslashbox{}{\\Potential\\Models}
&\multicolumn{2}{c|}{Cornell}&\multicolumn{2}{c|}
{Screened}&\multicolumn{2}{c|}{NR}&\multicolumn{2}{c|}
{BT}&\multicolumn{2}{c}{Fulcher}\\
\hline
k & $\langle r\rangle_{21}$ & $\Gamma$ & $\langle r\rangle_{21}$ & $\Gamma$ 
& $\langle r\rangle_{21}$ & $\Gamma$ & $\langle r\rangle_{21}$ & $\Gamma$ & 
$\langle r \rangle_{21}$ & $\Gamma$ \\
(MeV)&$({\rm GeV}^{-1})$&$({\rm keV})$&$({\rm GeV}^{-1})$&$({\rm
keV})$&$({\rm GeV}^{-1})$&$({\rm keV})$&$({\rm GeV}^{-1})$&$({\rm
keV})$&$({\rm GeV}^{-1})$&$({\rm keV})$\\
\hline
331&3.06&587&3.54&786&3.39&720&3.37&712&3.32&692\\
\hline
\end{tabular}
\end{table}

For the sake of completeness, 
finally we would like to contrast the $\eta_{c2}\to J/\psi(\psi^\prime)+\gamma$ 
processes with the dominant electromagnetic
transition of $\eta_{c2}$, the parity-changing electric transition $\eta_{c2}\to h_c +\gamma$. 
With the knowledge of the leading electric-dipole operator in (\ref{gammapNRQCD:Lag}) and 
the vaccum-to-$h_c$ matrix element in (\ref{wavefun:hc}), 
one can readily reproduce the well-known $E1$ transition formula:
\begin{subequations}
\bqa
& & \Gamma[\eta_{c2}\to h_c +\gamma] = {8\alpha e_c^2 \over 15}
\left| \langle r\rangle_{21} \right|^2,
\\
& & \langle r\rangle_{21} = \int^\infty_0 \!\!\! dr\: R_{^1P_1}(r) R_{^1D_2}(r)\,r^3.
\label{partial:width:etac2:hc:gamma}
\eqa
\end{subequations}
The contributions from the higher-order multipoles, $M2$ and $E3$, 
as well as the relativistic corrections, 
neither of which are expected to be significant, 
have been neglected for simplicity. 
From Table~\ref{table3}, one can see that the predicted partial width for 
$\eta_{c2}\to h_c +\gamma$ in five potential models ranges from $600$ to $800$ keV~\footnote{The $E1$ transition rates
tabulated in Table~\ref{table3} seem to be
somewhat larger than the often quoted $464$ keV~\cite{Barnes:2003vb}.}. 
As expected, this transition rate is several orders of magnitude greater than that of 
$\eta_{c2}\to J/\psi(\psi^\prime)+\gamma$.

In summary, we have presented a detailed analysis to the radiative
transitions $\eta_{c2}\to J/\psi(\psi^\prime)+\gamma$ within various
phenomenological potential models, employing the pNRQCD formalism as
an elegant and efficient calculational framework. The major
motivation of this study is to examine whether identifying the
$X(3872)$ with the $\eta_{c2}(1D)$ charmonium is compatible with the
$B$ factory measurements of the radiative decay  $X(3872)\to
J/\psi(\psi^\prime)+\gamma$. Our study reveals such an assignment
would cause severe contradictions with the available \textsc{BaBar}
measurements, either for the absolute branching fraction of $X\to \psi^\prime+\gamma$, 
or the ratio of this branching fraction to that of $X\to J/\psi+\gamma$. 
As a result, if the \textsc{BaBar} measurements~\cite{Aubert:2008rn} are
trustable, the possibility for the $X(3872)$ to be
identified with the $\eta_{c2}$ then becomes rather low. In our opinion, the
$X(3872)$ is most likely to carry $J^{PC}=1^{++}$.

A nuisance is that the preliminary \textsc{Belle} results~\cite{Bhardwaj:2010} on the
decay $X(3872)\to \psi'\gamma$ seem to conflict with the
\textsc{BaBar} measurement~\cite{Aubert:2008rn}. It is important and urgent to resolve
this experimental discrepancy, in order to unambiguously unravel the nature
of the X(3872).


\begin{acknowledgments}
We thank Juan Zhang for checking some of numerical
calculations.
This research was supported in part by the National Natural Science
Foundation of China under Grants No.~10875130 and No.~10935012.
\end{acknowledgments}



\begin{thebibliography}{99}

\bibitem{Brambilla:2004wf}
For a review, see  N.~Brambilla {\it et al.}  [Quarkonium Working
Group],
  arXiv:hep-ph/0412158;
\\
E.~S.~Swanson,
  Phys.\ Rept.\  {\bf 429}, 243 (2006)
  [arXiv:hep-ph/0601110];
\\
  N.~Drenska, R.~Faccini, F.~Piccinini, A.~Polosa, F.~Renga and C.~Sabelli,
  arXiv:1006.2741 [hep-ph].


\bibitem{Choi:2003ue}
  S.~K.~Choi {\it et al.}  [Belle Collaboration],
  Phys.\ Rev.\ Lett.\  {\bf 91}, 262001 (2003).

\bibitem{Aubert:2004ns}
  B.~Aubert {\it et al.}  [BABAR Collaboration],
  Phys.\ Rev.\  D {\bf 71}, 071103 (2005).


\bibitem{Acosta:2003zx}
  D.~E.~Acosta {\it et al.}  [CDF II Collaboration],
  Phys.\ Rev.\ Lett.\  {\bf 93}, 072001 (2004).

\bibitem{Abazov:2004kp}
  V.~M.~Abazov {\it et al.}  [D0 Collaboration],
  Phys.\ Rev.\ Lett.\  {\bf 93}, 162002 (2004).

\bibitem{Barnes:2003vb}
  T.~Barnes and S.~Godfrey,
  Phys.\ Rev.\  D {\bf 69}, 054008 (2004)
  [arXiv:hep-ph/0311162].


\bibitem{Tornqvist:2003na}
  N.~A.~Tornqvist,
  arXiv:hep-ph/0308277;
\\
  F.~E.~Close and P.~R.~Page,
  Phys.\ Lett.\  B {\bf 578}, 119 (2004);
\\
  M.~B.~Voloshin,
  Phys.\ Lett.\  B {\bf 579}, 316 (2004);
\\
  E.~Braaten and M.~Kusunoki,
  Phys.\ Rev.\  D {\bf 69}, 074005 (2004);
\\
  N.~A.~Tornqvist,
  Phys.\ Lett.\  B {\bf 590}, 209 (2004);
\\
  E.~S.~Swanson,
  Phys.\ Lett.\  B {\bf 598}, 197 (2004).


\bibitem{Maiani:2004vq}
  L.~Maiani, F.~Piccinini, A.~D.~Polosa and V.~Riquer,
  Phys.\ Rev.\  D {\bf 71}, 014028 (2005)
  [arXiv:hep-ph/0412098].

\bibitem{Li:2004sta}
  B.~A.~Li,
  Phys.\ Lett.\  B {\bf 605}, 306 (2005)
  [arXiv:hep-ph/0410264].

\bibitem{Abulencia:2005zc}
  A.~Abulencia {\it et al.}  [CDF Collaboration],
  Phys.\ Rev.\ Lett.\  {\bf 96}, 102002 (2006).

\bibitem{Aubert:2006aj}
  B.~Aubert {\it et al.}  [BABAR Collaboration],
  Phys.\ Rev.\  D {\bf 74}, 071101 (2006).

\bibitem{Abulencia:2006ma}
  A.~Abulencia {\it et al.}  [CDF Collaboration],
  Phys.\ Rev.\ Lett.\  {\bf 98}, 132002 (2007).

\bibitem{delAmoSanchez:2010jr}
  P.~del Amo Sanchez {\it et al.}  [BABAR Collaboration],
  Phys.\ Rev.\  D {\bf 82}, 011101 (2010)
  [arXiv:1005.5190 [hep-ex]].

\bibitem{Godfrey:2008nc}
  S.~Godfrey and S.~L.~Olsen,
  Ann.\ Rev.\ Nucl.\ Part.\ Sci.\  {\bf 58}, 51 (2008).

\bibitem{Aubert:2008rn}
  B.~Aubert {\it et al.}  [BABAR Collaboration],
  Phys.\ Rev.\ Lett.\  {\bf 102}, 132001 (2009).

\bibitem{DeFazio:2008xq}
  F.~De Fazio,
  Phys.\ Rev.\  D {\bf 79}, 054015 (2009)
  [arXiv:0812.0716 [hep-ph]].


\bibitem{Li:2009zu}
  B.~Q.~Li and K.~T.~Chao,
  Phys.\ Rev.\  D {\bf 79}, 094004 (2009)
  [arXiv:0903.5506 [hep-ph]].

\bibitem{Wang:2010ej}
  T.~H.~Wang and G.~L.~Wang,
  arXiv:1006.3363 [hep-ph].

\bibitem{Bhardwaj:2010}
  V.~Bhardwaj [Belle Collaboration], talk presented in the 7th
  International Workshop on Heavy Quarkonium, Fermilab, Chicago,
  May 18-21, 2010 [http://conferences.fnal.gov/QWG2010].

\bibitem{Aubert:2005vi}
  B.~Aubert {\it et al.}  [BABAR Collaboration],
  Phys.\ Rev.\ Lett.\  {\bf 96}, 052002 (2006).

\bibitem{Sebastian:1988kw}
  K.~J.~Sebastian, H.~Grotch and X.~Zhang,
  Phys.\ Rev.\  D {\bf 37}, 2549 (1988).

\bibitem{Sebastian:1994pu}
  K.~J.~Sebastian,
  Phys.\ Rev.\  D {\bf 49}, 3450 (1994).

\bibitem{Sebastian:1996cy}
  K.~J.~Sebastian and X.~G.~Zhang,
  Phys.\ Rev.\  D {\bf 55}, 225 (1997).


\bibitem{Brambilla:2005zw}
  N.~Brambilla, Y.~Jia and A.~Vairo,
  Phys.\ Rev.\  D {\bf 73}, 054005 (2006)
  [arXiv:hep-ph/0512369].

\bibitem{Pineda:1997bj}
  A.~Pineda and J.~Soto,
  Nucl.\ Phys.\ Proc.\ Suppl.\  {\bf 64}, 428 (1998);
\\
  N.~Brambilla, A.~Pineda, J.~Soto and A.~Vairo,
  Nucl.\ Phys.\  B {\bf 566}, 275 (2000).


\bibitem{Manohar:1997qy}
  A.~V.~Manohar,
  Phys.\ Rev.\  D {\bf 56}, 230 (1997)
  [arXiv:hep-ph/9701294].

\bibitem{Gromes:1984ma}
  D.~Gromes,
  Z.\ Phys.\  C {\bf 26}, 401 (1984).

\bibitem{Rosner:2001nm}
  J.~L.~Rosner,
  Phys.\ Rev.\  D {\bf 64}, 094002 (2001)
  [arXiv:hep-ph/0105327].

\bibitem{Grotch:1982bi}
  H.~Grotch and K.~J.~Sebastian,
  Phys.\ Rev.\  D {\bf 25}, 2944 (1982).

\bibitem{Jacob:1959at}
  M.~Jacob and G.~C.~Wick,
  Annals Phys.\  {\bf 7}, 404 (1959)
  [Annals Phys.\  {\bf 281}, 774 (2000)].

\bibitem{Martin:1970}
  A.~D.~Martin and T.~D.~Spearman,
  {\it Elementary Particle Theory},
North-Holland Publishing Company, 1970.


\bibitem{Karl:1975jp}
  G.~Karl, S.~Meshkov and J.~L.~Rosner,
  Phys.\ Rev.\  D {\bf 13}, 1203 (1976).

\bibitem{Karl:1980wm}
  G.~Karl, S.~Meshkov and J.~L.~Rosner,
  Phys.\ Rev.\ Lett.\  {\bf 45}, 215 (1980).

\bibitem{:2009pn}
  M.~Artuso {\it et al.}  [CLEO Collaboration],
  Phys.\ Rev.\  D {\bf 80}, 112003 (2009).

\bibitem{Amsler:2008zzb}
  C.~Amsler {\it et al.}  [Particle Data Group],
  Phys.\ Lett.\  B {\bf 667}, 1 (2008).

\bibitem{Eichten:1978tg}
E.~Eichten {\it et al.},
  Phys.\ Rev.\  D {\bf 17}, 3090 (1978)
  [Erratum-ibid.\  D {\bf 21}, 313 (1980)];
  Phys.\ Rev.\  D {\bf 21}, 203 (1980).

\bibitem{Barnes:2005pb}
  T.~Barnes, S.~Godfrey and E.~S.~Swanson,
  Phys.\ Rev.\  D {\bf 72}, 054026 (2005)
  [arXiv:hep-ph/0505002].

\bibitem{Buchmuller:1980su}
  W.~Buchmuller and S.~H.~H.~Tye,
  Phys.\ Rev.\  D {\bf 24}, 132 (1981).

\bibitem{Fulcher:1991dm}
  L.~P.~Fulcher,
  Phys.\ Rev.\  D {\bf 44}, 2079 (1991).

\bibitem{Jia:2009yg}
  Y.~Jia, J.~Xu and J.~Zhang,
  Phys.\ Rev.\  D {\bf 82}, 014008 (2010)
  [arXiv:0901.4021 [hep-ph]].

\end{thebibliography}
\end{document}